# Detection and Visualization of Differential Splicing in RNA-Seq data with JunctionSeq

Stephen W. Hartley[1,*], James C. Mullikin[1]

[1]Comparative Genomics Analysis Unit, Cancer Genetics and Comparative Genomics Branch, National Human Genome Research Institute, National Institutes of Health, Bethesda, MD 20892, USA.

*To whom correspondence should be addressed: Tel: 301-451-0277; Email: stephen.hartley@nih.gov

# ABSTRACT

Although RNA-Seq data provide unprecedented isoform-level expression information, detection of alternative isoform regulation (AIR) remains difficult, particularly when working with an incomplete transcript annotation. We introduce *JunctionSeq*, a new method that builds on the statistical techniques used by the well-established *DEXSeq* package to detect differential usage of both exonic regions and splice junctions. In particular, *JunctionSeq* is capable of detecting differentials in novel splice junctions without the need for an additional isoform assembly step, greatly improving performance when the available transcript annotation is flawed or incomplete. *JunctionSeq* also provides a powerful and streamlined visualization toolset that allows bioinformaticians to quickly and intuitively interpret their results. We tested our method on publicly available data from several experiments performed on the rat pineal gland and *Toxoplasma gondii*, successfully detecting known and previously validated AIR genes in 19 out of 19 gene-level hypothesis tests. Due to its ability to query novel splice sites, *JunctionSeq* is still able to detect these differentials even when all alternative isoforms for these genes were not included in the transcript annotation. *JunctionSeq* thus provides a powerful method for detecting alternative isoform regulation even with low-quality annotations. An implementation of *JunctionSeq* is available as an R/Bioconductor package.

# INTRODUCTION

In 2015 alone, hundreds of research papers have reported differential gene expression (DGE) based on RNA-Seq data (1-10). In general, RNA-Seq studies focus primarily on detecting gene-wide effects, in which entire genes are upregulated or downregulated depending on some experimental or biological condition. Although the statistical methodologies have advanced considerably, these studies generally follow the same basic design principles as previous microarray-based expression research.

However, RNA-Seq data provide more than simple measurements of gene-level expression. In theory, RNA-Seq can be used to study more complex regulatory phenomena at the isoform level, even when the isoforms in question are unannotated. Numerous tools have been developed to detect alternative isoform regulation (AIR) (11); however, only a few of these tools have seen serious application outside their respective methodology papers, and many RNA-Seq studies do not even attempt the detection of AIR (12-14).

Detecting alternative isoform regulation is inherently difficult in RNA-Seq, as sequencer reads are often one or more orders of magnitude shorter than the transcripts themselves. While there are several utilities that attempt to de-convolute read data into isoform abundances, the accuracy and robustness of these methods is difficult to establish (15,16). Isoform expression estimates seem to vary considerably between different tools, and generally depend on the quality and completeness of the transcript assembly (17,18). Most of the newest and most popular tools do not assess or model unannotated isoforms (including *Kallisto* (19), eXpress (20), and *RSEM* (21)), and the presence of such isoforms can substantially alter the estimated abundances of the known isoforms belonging to a gene. Thus: accurate analysis of differential isoform regulation may be very difficult when the available transcript annotation is incomplete.

In addition, almost all existing AIR analysis tools share one common flaw: the results are difficult to interpret. This is not a trivial issue: unlike simple DGE, instances of AIR cannot be adequately characterized by a single fold change and p-value. Alternative isoform regulation is a broad and diverse class of phenomena that can involve alternative splice sites, alternative promoters, nucleosome occupancy, cassette exons, alternative donors/acceptors, long non coding RNAs, alternative polyadenylation, gene-level differential expression, or any number of these factors in combination. A gene may be composed of dozens of distinct isoforms, each controlled by its own set of regulatory mechanisms. As a consequence, the raw results of alternative isoform regulation analyses are often counterintuitive and resistant to interpretation.

The interpretation of these results is critical: in order to be considered credible by the community, any detected instances of AIR will generally require validation by secondary methods (such as qRT-PCR or SMRT sequencing). Such validation is often costly and time consuming. Detecting the mere presence of an effect is insufficient for these purposes: the investigator must also be able to identify which specific isoforms are being differentially used and assess the strength, direction, and credibility of the effect. Furthermore: since hundreds of AIR genes may be detected, this interpretation process must be streamlined, scalable, and intuitive.

The *DEXSeq* software package tests for differential usage of exonic regions as a proxy for differential isoform usage, and provides a powerful suite of visualization tools (22,23). However, *DEXSeq* is only effective at detecting AIR when it results in changes in the expression of the annotated exonic regions. This method has two major weaknesses: firstly, it does not query all forms of alternative isoform regulation, as not all forms of AIR necessarily produce differentials in the exon counts. Secondly, this method is strongly dependent on the reference annotation, and cannot directly identify differentials in novel exons and splice junctions. Additionally, *DEXSeq* output plots are often difficult to read and interpret, particularly for genes with a large number of isoforms and splice variants.

Here, we introduce *JunctionSeq*, a new method and associated Bioconductor package that builds upon the popular and well-established *DEXSeq* methodology in order to detect differential usage of both exons and known or novel splice junctions. Unlike most similar tools, *JunctionSeq* can reliably detect alternative isoform regulation even when the alternatively regulated isoforms themselves are not annotated. *JunctionSeq* also provides improved visualization tools that produce readable and informative expression profiles across all potential genes of interest, and across the genome as a whole.

# METHODS

## Differential Usage

Estimating the true isoform abundance of overlapping multi-kb transcripts using hundred-base-pair reads is an inherently difficult and error-prone task, particularly when some of the isoforms are not known *a priori*. Rather than attempting to directly estimate these isoform abundances, we instead attempt to detect differentials in quantities that are directly observable: the read counts for exonic regions and splice junction loci. Since all major aligners designed for RNA-Seq will align across novel (unannotated) splice junctions (24), we can also test for differential usage of these novel splicing variants.

"Differential usage" (DU) is an observed phenomenon in which individual transcript sub-components (in this case, exons or splice junctions) display expression that is inconsistent with that of the gene as a whole. This can sometimes be counterintuitive: if a gene is differentially expressed, an individual sub-component that displays constant expression across all samples might be considered "differentially used", as its expression is not consistent with that of the gene.

Testing for differential usage of splice junctions has a number of benefits. Firstly, since splice events are discrete and identifiable we can include novel splicing variants when they splice to/from known genes. This allows us to indirectly query for differential regulation in unknown isoforms, improving performance on sparsely annotated genomes. Furthermore, some forms of AIR do not necessarily result in observable differentials in the exon-level counts. An intron retention, for example, will alter splice junction counts but not the counts of the flanking exons. As a result, our method substantially broadens the variety of regulatory phenomena that can be effectively detected.

**Statistical Methodology**

Like the *DEXSeq* Bioconductor package, we first partition each gene into a set of mutually non-overlapping exonic regions, and then use the read (or read-pair) counts for each exonic region to estimate the relative expression of each exon for each experimental condition (22). Unlike the DEXSeq package, we also calculate counts for each splice junction belonging to each gene, including novel splice junctions that are within the gene's span that surpass a user-specified normalized mean coverage threshold (we recommend 3 reads per sample). We use the *DESeq2* package along with a set of specialized multivariate generalized linear models (GLM) to individually test for differential usage of each exonic region and splice junction (23).

Previous studies have done similar analyses simply by plugging splice junction counts into DEXSeq (25), however we found this method to be inadequate as it did not account for the numerous differences in the distribution and structure of the splice junction count data. A number of modifications to the basic *DEXSeq* methodology were found to be necessary.

To begin with: for most datasets DEXSeq will double- or triple-counts the vast majority of reads, as it uses the sum of all exonic regions as a proxy for estimating gene-level expression. This would be even more pronounced in *JunctionSeq*, and while it would not technically invalidate the hypothesis tests it can bias the fold-change estimates by over-weighting variant-dense regions, producing confusing artifacts under certain conditions. Thus, we use gene-level counts as the basis for our estimates of gene-wide expression rather than the sum of all exonic regions. This means that in the *JunctionSeq* framework no read or read-pair is counted more than once in any given statistical model. This model framework was applied to both the hypothesis test and the effect estimation steps. For similar reasons, our size factor estimation is carried out using the gene-level counts rather than the exon/junction counts.

In addition, we found that splice junctions and exonic regions sometimes followed different dispersion trends from one another. This is not surprising, given the various biological and technical differences between the two count types. To account for this difference, *JunctionSeq* (by default) fits separate dispersion trends for exonic regions and for splice junctions. As in *DEXSeq*, the final dispersion estimates used for hypothesis testing are calculating by estimating the maximum *a priori* (MAP) dispersions for each exon and junction, which "shrinks" each feature-specific dispersion estimate towards its respective fitted dispersion estimate (23).

For a complete description of the *JunctionSeq* methodology, see Section 3 of the online supplement.

**The Interpretation Problem**

Most existing AIR utilities provide little-to-no functionality to assist the end-user in the interpretation of the results. Some tools provide basic analysis-wide summary plots (22,26,27) or expression profile plots for individual samples (13,27-32), but very few provide methods for directly comparing gene expression profiles between multi-sample experimental groups. Many tools provide little information to the user beyond p-values and fold changes (11).

The *DEXSeq* visualization toolset, while unparalleled in its class, was found to be insufficient for our purposes (22). *DEXSeq* generates a number of gene profile plots that show read/read-pair coverage across each exonic region, plotted above a representation of the isoform annotation (see Figure 2b). However, genes vary widely in the number of exons and isoforms they possess, and as a result these plots vary widely in the complexity of the data they present. Consequently: regardless of the specific graphical settings, *DEXSeq*-generated plots often suffer from over-plotting, severely reducing their utility.

*JunctionSeq* implements a number of refinements designed to streamline and improve this process. Many parameters are automatically adjusted for each figure to improve readability, including adjustments to the feature label size and orientation, figure aspect ratio, relative size of the left and right panels, y-axis scaling, figure margins, and label positioning (See Figure 1, Figure 2, and Figure 4). Other improvements were added to make the plots more informative, including the nonlinear expansion of small features, highlighting of significant features, nested splice junction diagrams, and the inclusion of a gene-level expression plot. The various plots can either be viewed manually or browsed using a set of automatically-generated html pages, designed for easy navigation between genes and between experiments.

While these features might seem cosmetic, they vastly improve the utility and scalability of this tool and allow investigators to quickly examine a large number of potential genes of interest in order to identify, characterize, and assess interesting biological phenomena.

The *JunctionSeq* analysis pipeline also generates genome-wide browser tracks suitable for use with IGV or the UCSC genome browser (See Figure 3). These tracks allow investigators to interactively browse expression profiles, splice junction counts, and statistically significant features across the entire genome, all alongside the numerous publicly available annotation tracks (32,33).

# RESULTS

To demonstrate the strengths of our new method, we applied JunctionSeq to two different publicly available datasets. The first was in *Toxoplasma gondii* and included 3 analyses; the second was in rat pineal glands and included 4 analyses. Both datasets included known and validated AIR genes, one gene in the *Toxoplasma gondii* dataset and four genes in the rat pineal gland dataset. Thus, there were a total of 19 gene-level hypothesis tests in which we expected to detect differential usage, acting as positive controls in our analysis.

*JunctionSeq* consistently detected differential usage in all known AIR genes across all experiments, even when the alternative isoforms were not included in the transcript annotation.

**Test dataset 1: *Toxoplasma gondii* and TgSR3**

Our first test dataset originated from a previous study in which alternative splicing was detected and validated in *Toxoplasma gondii* between control samples and samples in which overexpression of the TgSR3 gene was induced (34). There were four sample groups of 3 biological replicates each: untreated; induced, 4 hours; induced, 8 hours; and induced, 24 hours. The dataset is available from the NCBI short read archive (SRA), accession number PRJNA252680.

In the original study numerous genes were found to display differential splicing between the induced and untreated sample groups. One particular gene, TGGT1_207900, was found to display strong differential splicing across an unannotated 5' variant in all three comparisons. This effect was detected using *DEXSeq* via a *CuffLinks* assembly, and was subsequently confirmed via qRT-PCR. In order to demonstrate JunctionSeq's ability to detect differential usage of novel variants, we performed the same analysis using *JunctionSeq*, but without the benefit of the *CuffLinks* assembly step.

*Detection of AIR without CuffLinks assembly:*

Even without a complete transcript assembly, *JunctionSeq* detected differential usage of the previously-validated novel splice variant in TGGT1_207900 in all three experiments, with adjusted p-values of 0.00023, $8.5 \times 10^{-13}$, and 0.0098 for the untreated vs 4-hour, 8-hour, and 24-hour experiments, respectively. The gene profile plots clearly displayed the same form of differential splicing found in the original experiment (see Figure 1 and Supplemental figs. 1-2) (34).

This demonstrates that *JunctionSeq* can accurately detect differentials in novel splicing variants, and does not require a complete and comprehensive transcript annotation in order to detect alternative isoform regulation.

**Test dataset 2: Circadian Rhythms in the Rat Pineal Gland**

The rat pineal gland is known to display strong and consistent differential expression resulting from neural stimulation across hundreds or thousands of genes (35,36). Most if not all of these changes are believed to be controlled via neural innervation of the pineal gland by the SCG, using the neurotransmitter norepinephrine (NE) and the second messenger cyclic AMP (cAMP) (37-44).

Several genes have already been found in the literature to exhibit neurally-controlled alternative isoform usage in the rat pineal gland: Crem (45-47), Pde4b (48), Atp7b (49), and Slc15a1 (formerly known as Pept1) (50,51). It should be noted that all of these genes were discovered in previous studies using different datasets, and all are validated and well-established in the literature.

We performed four comparisons in which we expected to detect differential splicing in genes that are neurally controlled by norepinephrine and cAMP: two *in vivo* analyses comparing night and day conditions in no-surgery (Ctrl) and sham-surgery (Sham) rats, as well as two *in vitro* analyses comparing pineal glands in organ culture that had been treated with norepinephrine (NE) or dibutyryl cyclic AMP (DBcAMP, an analogue of the second messenger, cyclic AMP), each versus an untreated control set (CN). Given that the four known-AIR genes are neurally controlled, we expect to detect differential usage in all four genes across all four comparisons.

*Detection of Known AIR genes:*

For the four known AIR genes, we found strong genome-wide statistical significance for all 16 gene-level hypothesis tests (see Table 2). The genes Crem, Pde4b, and Atp7b were detected by *JunctionSeq* at an extremely high significance level in all analyses (p-adjust < 1e-8 for all three genes and all four comparisons), and the gene Slc15a1 was detected at a moderately high significance level in all analyses (p-adjust < 0.01). See Figure 2a for an example plot displaying the *JunctionSeq* results for the Crem gene in the sham-surgery group.

*Differential usage of novel variants*

To demonstrate *JunctionSeq's* ability to detect differential usage of novel splice junctions even with an incomplete transcript assembly, we performed a second set of analyses with a reduced annotation. For each of the three known AIR genes that had multiple annotated transcripts (Crem, Pde4b, and Atp7b), we manually removed all but one transcript from the ensembl annotation GTF and then re-ran the analyses. This was intended to simulate the scenario in which AIR occurs in poorly-annotated genes. The gene Slc15a1 only has one transcript in the current annotation, and thus the annotation was left unchanged.

Even with only one annotated transcript, *JunctionSeq* was still able to detect differential usage of "novel" splice sites for all four genes across all four comparisons (p-adjust < 0.01, see the right half of Table 2). See Figure 4a for an example plot displaying the incomplete-annotation *JunctionSeq* results for the Crem gene in the sham-surgery group.

*Replicability and consistency:*

In addition to confirming known AIR genes and providing a strong positive control for *JunctionSeq*, we can further use these analyses to demonstrate the reliability and replicability of our methods by examining the overlap between the four comparisons.

While these experiments are not direct replications, isoforms whose regulation is controlled specifically by neural innervation of the pineal gland through the SCG (via norepinephrine and cyclic AMP) should theoretically exhibit similar expression regulation across all four experiments.

In each comparison hundreds of genes were found to display statistically significant differential exon or splice-junction usage (at p-adjust < 0.01), and 42 of these genes displayed differential usage in all four analyses (see Table 1 and Supplemental fig. 3). The strong concordance between the four experiments spanning very different (but biologically related) experimental conditions demonstrates that *JunctionSeq* produces consistent and replicable results.

**Comparison with Existing Tools**

Comparisons between differential isoform regulation tools are difficult, as many are actually designed to detect subtly distinct phenomena. As a consequence: even if both tools perform with perfect accuracy they may still return different results. *CuffDiff*, for example, performs alternative isoform regulation analysis that segregates genes by promoter site, excluding detection of differential alternative promotor usage. This would exclude all five of the previously-validated cases of alternative isoform usage found in our test datasets. Other transcript-alignment-based tools like *Kallisto/Sleuth* (19), *eXpress* (20), and *RSEM* (21) only detect overall differential expression of individual transcripts and do not attempt to detect differential usage of transcripts relative to one another. Thus, results from these tools would likely consist predominantly of differentially expressed genes, and would not specifically target differential splicing. Furthermore, most such tools are strongly annotation-dependent and do not attempt to assess novel splice variants.

The obvious comparison, however, is with the *DEXSeq* software tool (22). We found that when the affected isoforms are known, *JunctionSeq* appears to perform at least as well (or better) than *DEXSeq*. However, when unannotated isoforms are involved *JunctionSeq* demonstrates clear superiority due to its ability to query unannotated splice junctions.

*Toxoplasma gondii analyses*

Without a CuffLinks assembly, *DEXSeq* was unable to detect any differential usage in the validated gene (TGGT1_207900) in any of the three *Toxoplasma gondii* analyses (see Figure 1b and Supplemental figures 1b and 2b). *JunctionSeq*, however, detects the differential usage of the alternative start site in all three analyses, even without the *CuffLinks* assembly.

*Rat pineal gland analyses*

In the rat pineal gland data, *JunctionSeq* and *DEXSeq* seemed to perform similarly when the full transcript annotation was used (see Table 1, Supplementary Table 1, and Supplementary Figures 3-4). Across all experiments *JunctionSeq* detected at least as many statistically significant genes in each experiment individually and found more genes that overlapped between all four analyses.

For the four known-AIR genes, *DEXSeq* and *JunctionSeq* return very similar results when the full annotation was used, although *JunctionSeq* reported slightly weaker significance for the gene Slc15a1 (see Table 2 and Table 3).

When the transcript annotation was incomplete, *DEXSeq* fails to detect differential usage in 3 of the 16 tests, one for Crem and two for Pde4b (at p-adjust < 0.01, see Table 3). JunctionSeq, on the other hand, still reports differential usage in all 16 tests (see Table 2). Furthermore, although the other five *DEXSeq* tests for the genes Crem and Pde4b are still statistically significant at p-adjust < 0.01, all of the reported p-values are several orders of magnitude weaker than those found in either the corresponding *JunctionSeq* analyses or the corresponding full-annotation *DEXSeq* analyses.

Both the rat pineal gland and *Toxoplasma gondii* analyses lead us to the same conclusions: when the transcript annotation is complete and comprehensive, *DEXSeq* and *JunctionSeq* produce very similar results. However, when novel isoforms are involved, *JunctionSeq* provides superior performance.

*Other Advantages of JunctionSeq*

The improved visualization tools provided by *JunctionSeq* further increase its utility. In general, simply detecting the presence of AIR is insufficient; the investigator must also be able to determine precisely which isoforms or splice variants are responsible for the apparent differentials. In many cases

*DEXSeq* detects differentials in the same genes as *JunctionSeq*, however, even when manually examining the DEXSeq plots it is often impossible to identify the specific splice variants that are being differentially expressed, particularly when the relevant exons or splice junctions are unannotated.

For example, in the Crem gene there are several clusters of small exonic regions (E009-E010, E013-E015, E017-E021, see Figure 2) that are indistinguishable in the *DEXSeq* gene/transcript diagram (bottom of Figure 2b), but can be easily identified and matched to their corresponding isoforms in the JunctionSeq plot (bottom of Figure 2a and c), due to the nonlinear expansion of small features.

Similarly, when novel isoforms are involved, it is often impossible to identify the relevant splicing variants in the DEXSeq plots, even when statistical significance is detected in the gene. This is because DEXSeq will often detect the indirect effects of alternative isoform usage, but the causal variants themselves will remain obscured.

For example, in the incomplete-annotation analysis shown in Figure 4b, the first exon (E001) actually displays borderline statistical significance in the *DEXSeq* analysis (p-adjust = 0.016), due to the fact that this exon is not present in the (unobserved) alternative isoforms. However, even if this is considered significant it is impossible to identify the actual variants responsible for this effect, as they are not directly observable in the *DEXSeq* plots. The *JunctionSeq* plots, on the other hand, clearly show the source of the differential usage in the various "novel" splice junctions, most of which lead to the upstream alternative promoter site.

If desired, *JunctionSeq* can (optionally) run pure exon-based analyses, reducing the number of comparisons. One of the major strengths of *JunctionSeq* is that it queries a broader array of regulatory phenomena, however this comes at the cost of additional comparisons, potentially reducing power. Running a purely exon-based analysis may provide superior results when working with a well-characterized tissue on a comprehensively annotated genome. In fact, with a certain set of options (documented in the user manual), *JunctionSeq* will precisely reproduce a standard *DEXSeq* analysis while still providing the user with the enhanced visualization tools of *JunctionSeq*. To demonstrate the advantages of the *JunctionSeq* plotting engine we plotted identical analyses run by *JunctionSeq* and *DEXSeq* for a large and complex human gene using simulated data (see Supplemental figures 8-9).

# DISCUSSION

*JunctionSeq* offers a powerful, flexible, statistically robust, and efficient solution for the identification, characterization, and interpretation of differential isoform regulation. The underlying methodology has a strong theoretical basis and is built upon established statistical methods that are already widely accepted by the community. It includes a number of powerful improvements that allow it to query a broader class of regulatory phenomena, including the differential regulation of novel splicing variants in the absence of an accurate and comprehensive transcript annotation.

This is a critical addition to the community, as DEXSeq cannot consistently detect differentials in novel variants, and many popular tools such as *Kallisto* (19), eXpress (20), or *RSEM* (21) cannot assess novel variants at all. Furthermore, many transcript quantification tools seem to perform poorly when used with an incomplete transcript annotation (17). Although *JunctionSeq* may not necessarily provide uniform superiority over existing methods when the annotation is comprehensive, it provides a valuable tool for researchers studying esoteric tissues and/or less-common species.

Another critical advantage of the *JunctionSeq* software toolset is its suite of powerful automated visualization and interpretation tools, which allow investigators to quickly and intuitively examine hundreds of genes. This assists investigators in identifying and characterizing genes of interest for further validation and study.

**Example Interpretation**

For the purposes of demonstration we will examine a well-known AIR gene, Crem, in the rat pineal gland dataset. The mechanism behind the circadian alternative isoform regulation of the Crem gene is already well understood, and the patterns of expression of this gene's various isoforms are well-characterized (45,46,52). Briefly: an internal promoter is greatly upregulated at night, resulting in large quantities of a number of small transcripts collectively known as ICER. ICER is known to play a major role in the melatonin synthesis pathway (53).

By default, *JunctionSeq* automatically generates gene profile plots for every gene that contains one or more differentially used exon or splice junction. Figure 2 (a, c, and d) displays a few of the available plots in the sham-surgery night/day experiment.

As seen in the small rightmost panel of each *JunctionSeq* plot (i.e. the narrow panels labelled "GENE"), the Crem gene as a whole appears to display strong upregulation at night (~15,000 vs ~650 read-pairs per sample). Looking at the gene profile plots we can see that this is not uniform across the gene: some of the exonic regions and splice junctions display strong upregulation at night while others do not. Exonic regions E009 through E021 all display strong differentials (>8x, see Figure 2c), but exonic regions E001 through E008 display consistently low counts at both day and night. The splice junction plot (see Figure 2d) shows similar results for the splice junction coverage.

It may seem counterintuitive that the constant-expression exons (E001 through E008) are marked as statistically significant. This is because *JunctionSeq* (like *DEXSeq*) tests for differential usage, not differential expression. The expression of each sub-feature is compared with the expression of the gene as a whole (see the rightmost panel of each *JunctionSeq* plot in Figure 2). Since the gene as a whole has a strong differential, exonic regions and splice junctions that do **not** display such differentials are considered "differentially used" relative to the gene.

Using the genome browser tracks produced in the *QoRTs*/*JunctionSeq* pipeline (Figure 3), we can examine the read coverage across the genome and over all known and novel splice junctions. These can be examined alongside external annotation tracks such as the RepeatMasker track or the UCSC-maintained EST and mRNA databases. Visual examination alongside these tracks can be critical, as it can determine whether novel splice variants have been previously detected, or if apparent differentials might be the result of alignment artifacts or flaws in the annotation.

Taken together, these visualizations lead towards a clear and obvious hypothesis: the full-length isoforms of the Crem gene display constant low-level expression at day and night, whereas the isoforms originating in the internal ("ICER") promoter are greatly upregulated at night. This "hypothesis" matches the known behavior and function of this gene in the literature (45,46,52,53).

Similar plots are available online for the Crem gene in the other three rat pineal gland experiments (See Supplemental Figures 5-7).

**The *JunctionSeq* R package**

We implemented the described method in a new Bioconductor package, *JunctionSeq*, written entirely in the R statistical programming language.

The *JunctionSeq* analysis pipeline requires the *QoRTs* quality-control/data-processing software package (54) in order to generate the raw gene, exon, and splice junction counts. *QoRTs* is also used to create the multi-sample normalized-mean "wiggle" tracks for use with IGV or the UCSC genome browser.

The *JunctionSeq* package is extensively documented and includes a comprehensive walkthrough and example dataset, with line-by-line instructions describing the complete analysis pipeline. *JunctionSeq* will be included in Bioconductor release 3.3 (http://bioconductor.org/packages/JunctionSeq/), and is available now along with additional online help and documentation at the *JunctionSeq* webpage: http://hartleys.github.io/JunctionSeq/.

# FUNDING

This research was supported by the Intramural Research Program of the National Human Genome Research Institute, National Institutes of Health.

# DATA ACCESS

The datasets used in the application sections are available from the NCBI short read archive (SRA), with accession numbers PRJNA267246 and PRJNA252680 for the *Rattus norvegicus* (36) and the *Toxoplasma gondii* (34) datasets, respectively.

# ACKNOWLEDGEMENTS:

We would like to thank Steven L. Coon and David C. Klein from the Eunice Kennedy Shriver National Institute of Child Health and Human Development for their input and assistance, particularly in generating and conceiving the rat pineal experiments.

We would also like to thank John Didion and Peter Chines from the Medical Genomics and Metabolic Genetics Branch at the National Human Genome Research Institute for their assistance in the testing and development of this software.

# TABLES

**Table 1: *JunctionSeq* results.** The numbers of genes found to exhibit significant differential exon or splice junction usage for the four rat pineal gland analyses at various p-value thresholds. A similar table for the *DEXSeq* analyses is available online.

| Adjusted p-value threshold | *In vivo* | | | *In vitro* | | | Overlap, All Four |
|---|---|---|---|---|---|---|---|
| | Ctrl Day/Night | Sham Day/Night | Overlap, *in vivo* | CN vs NE | CN vs DBcAMP | Overlap, *in vitro* | |
| 0.01 | 447 | 320 | 168 | 144 | 195 | 90 | 42 |
| 0.001 | 300 | 202 | 116 | 89 | 127 | 61 | 28 |
| 0.0001 | 227 | 151 | 94 | 67 | 91 | 48 | 18 |
| 0.00001 | 182 | 119 | 79 | 51 | 74 | 38 | 15 |
| 0.000001 | 151 | 98 | 61 | 43 | 65 | 34 | 14 |

**Table 2:** JunctionSeq gene-level adjusted p-values for 4 known-AIR genes in the rat pineal gland, both with and without a complete isoform annotation. The left 4 columns display the results from a normal analysis, the right 4 columns display the results from an analysis in which all but one isoform was removed from the annotation for each gene, simulating a scenario in which the gene is poorly studied and the annotation incomplete. *Note: since the Slc15a1 gene actually only has one known transcript, the "full" and "incomplete" annotation analyses for this gene are equivalent, differing only slightly due to minor analysis-wide differences in the dispersion estimation and multiplicity correction.

| Gene Symbol | Full Annotation | | | | Incomplete Annotation (1 "known" isoform) | | | |
|---|---|---|---|---|---|---|---|---|
| | Ctrl Day/Night | Sham Day/Night | CN vs NE | CN vs DBcAMP | Ctrl Day/Night | Sham Day/Night | CN vs NE | CN vs DBcAMP |
| Atp7b | *<1e-8* | *<1e-8* | *<1e-8* | *<1e-8* | *<1e-8* | *<1e-8* | *<1e-8* | *<1e-8* |
| Crem | *<1e-8* | *<1e-8* | *<1e-8* | *<1e-8* | *<1e-8* | *<1e-8* | *<1e-8* | *<1e-8* |
| Pde4b | *<1e-8* | *<1e-8* | *<1e-8* | *<1e-8* | *<1e-8* | *<1e-8* | *<1e-8* | *2.8e-7* |
| Slc15a1 | *<1e-8* | *<1e-8* | *0.0034* | *0.0015* | *<1e-8** | *<1e-8** | *0.0034** | *0.0016** |

**Table 3:** DEXSeq gene-level adjusted p-values for 4 known-AIR genes in the rat pineal gland, both with and without a complete isoform annotation. See Table 2. Note that without the complete annotation, several tests do not show significant differential usage or have much less significant p-values.

| Gene Symbol | Full Annotation | | | | Incomplete Annotation (1 "known" isoform) | | | |
|---|---|---|---|---|---|---|---|---|
| | Ctrl Day/Night | Sham Day/Night | CN vs NE | CN vs DBcAMP | Ctrl Day/Night | Sham Day/Night | CN vs NE | CN vs DBcAMP |
| Atp7b | *<1e-8* | *<1e-8* | *<1e-8* | *<1e-8* | *<1e-8* | *<1e-8* | *<1e-8* | *<1e-8* |
| Crem | *<1e-8* | *<1e-8* | *<1e-8* | *<1e-8* | *1.2e-5* | *0.0357* | *6.5e-6* | *9.3e-5* |
| Pde4b | *<1e-8* | *<1e-8* | *<1e-8* | *<1e-8* | *1.2e-4* | *0.0041* | *1.0* | *1.0* |
| Slc15a1 | *<1e-8* | *<1e-8* | *3.3e-5* | *2.7e-5* | *<1e-8** | *<1e-8** | *3.2e-5** | *3.0e-5** |

# FIGURES

**Figure 1:** Gene profile plots from (a) JunctionSeq and (b) DEXSeq for the TGGT1_207900 gene, in the 8-hour-induced vs un-induced *Toxoplasma gondii* experiment. The large central plotting panel of (a) and (b) displays the estimates for the mean normalized read counts for each exon or splice junction for the 8-hour-induced (red) or uninduced (blue) sample groups. The narrow panel on the right in (a) displays the gene-level mean normalized read counts. In each plot a gene diagram is drawn beneath the main plotting panels, showing the location and layout of the gene. Statistically significant (p-adjust < 0.01) exons or junctions are drawn with pink, and features that had counts that were too low to test are drawn in light gray (or they would be, if there were any such features). Known splice junctions are drawn with solid lines and unannotated splice junctions are drawn with dashed lines. Note in the JunctionSeq plot the first two splice junctions are strongly and significantly differentially used (in opposing directions). This effect was confirmed in a previous study via qRT-PCR (34). Also note that differential usage is not apparent in the DEXSeq plot, as the differentially used features are unannotated. Similar plots for the other two *Toxoplasma gondii* experiments can be found online, and show similar results (see supplemental figs. 1-2).

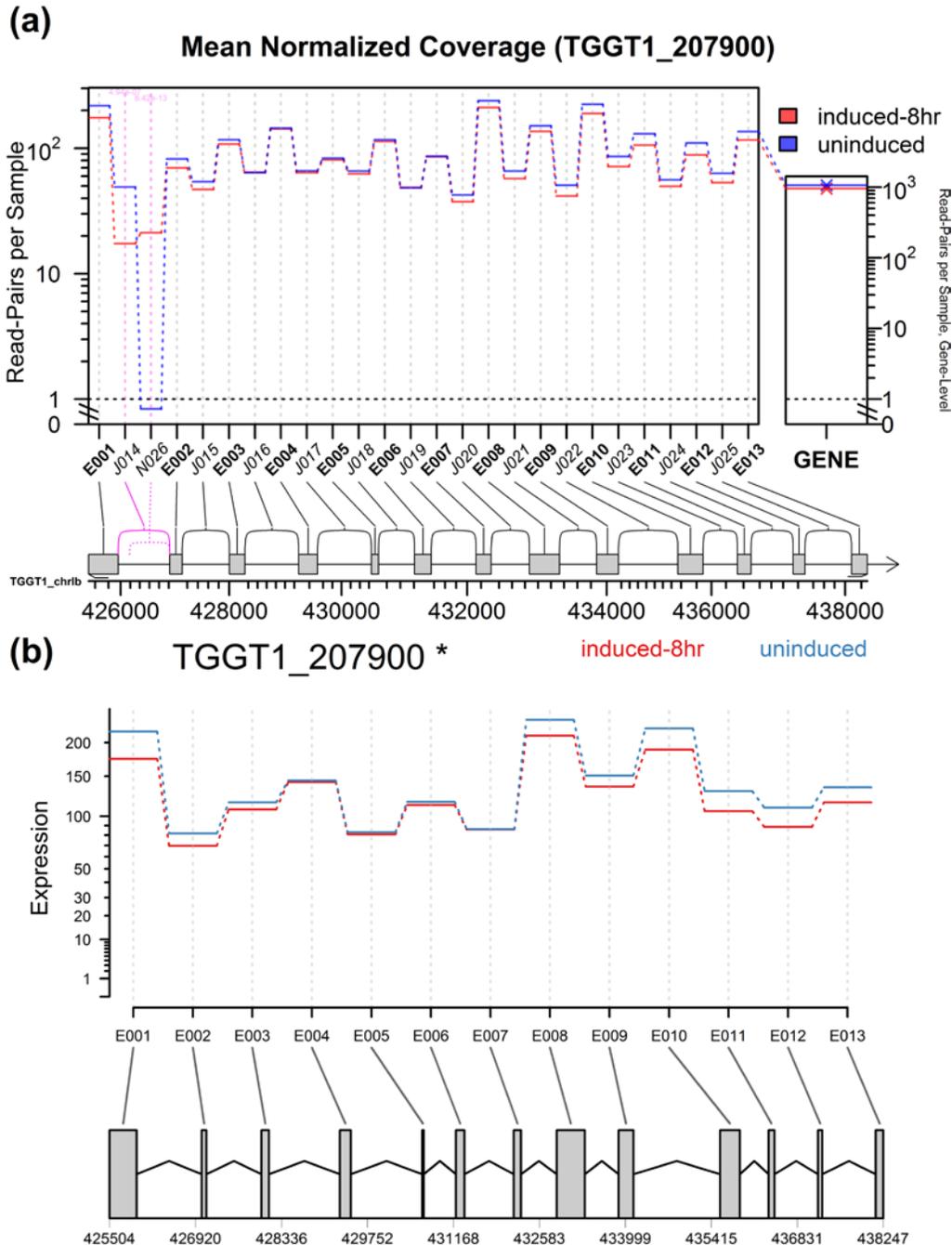

**Figure 2:** Day/Night gene profile plots for the Crem gene in the rat pineal gland, sham-surgery group. Plots (a), (c), and (d) were produced by JunctionSeq, and (b) was produced by an equivalent analysis using DEXSeq. The full standard JunctionSeq gene profile plot (a) includes both exon and splice junction information. The equivalent DEXSeq plot (b) only displays exon information. Optionally, JunctionSeq can produce similar exon profile plots (c), or plots displaying only splice-junction information (d). Beneath the plotting regions in each figure a gene diagram displays the features' positions on the genomic scale (note that small features are expanded for readability in the JunctionSeq versions). Novel junction loci are drawn using dashed lines. In the upper plots (a and b), all known transcripts are displayed beneath the main plotting area. Similar plots are available online for the control day/night comparison, as well as the two treated-vs-untreated comparisons (see Supplemental Figures 5-7).

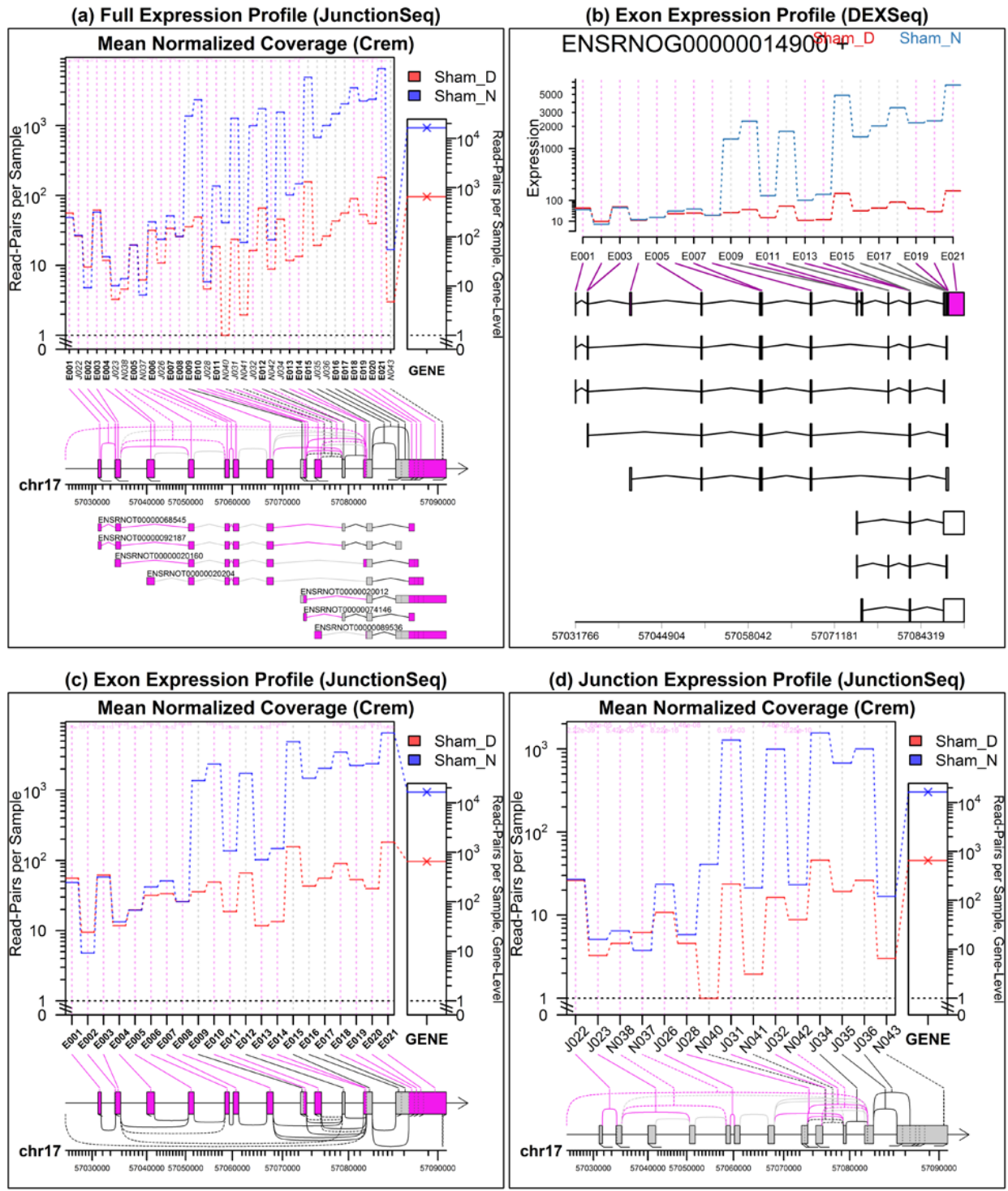

**Figure 3:** Genome-wide Browser tracks produced in the *QoRTs/JunctionSeq* pipeline. The above screenshot displays much of the same information found in Figure **2**, except using the UCSC genome browser. The top track displays the ensembl gene annotation. The second track displays the statistically significant features, with the adjusted p-value included in parentheses. The next track is a "wiggle" track that displays coverage over both the forward and reverse strand (above and below the x-axis, respectively), in red and blue for day and night, respectively (overlap is colored black). The next two tracks display all exons and splice junctions, respectively, that were tested for DU by *JunctionSeq*. The day/night normalized mean expression values from Figure **2** are included in parentheses. The final track is from RepeatMasker, and displays regions with repeating or low-complexity elements. Using these tracks together can be vital for the purposes of interpretation and validation.

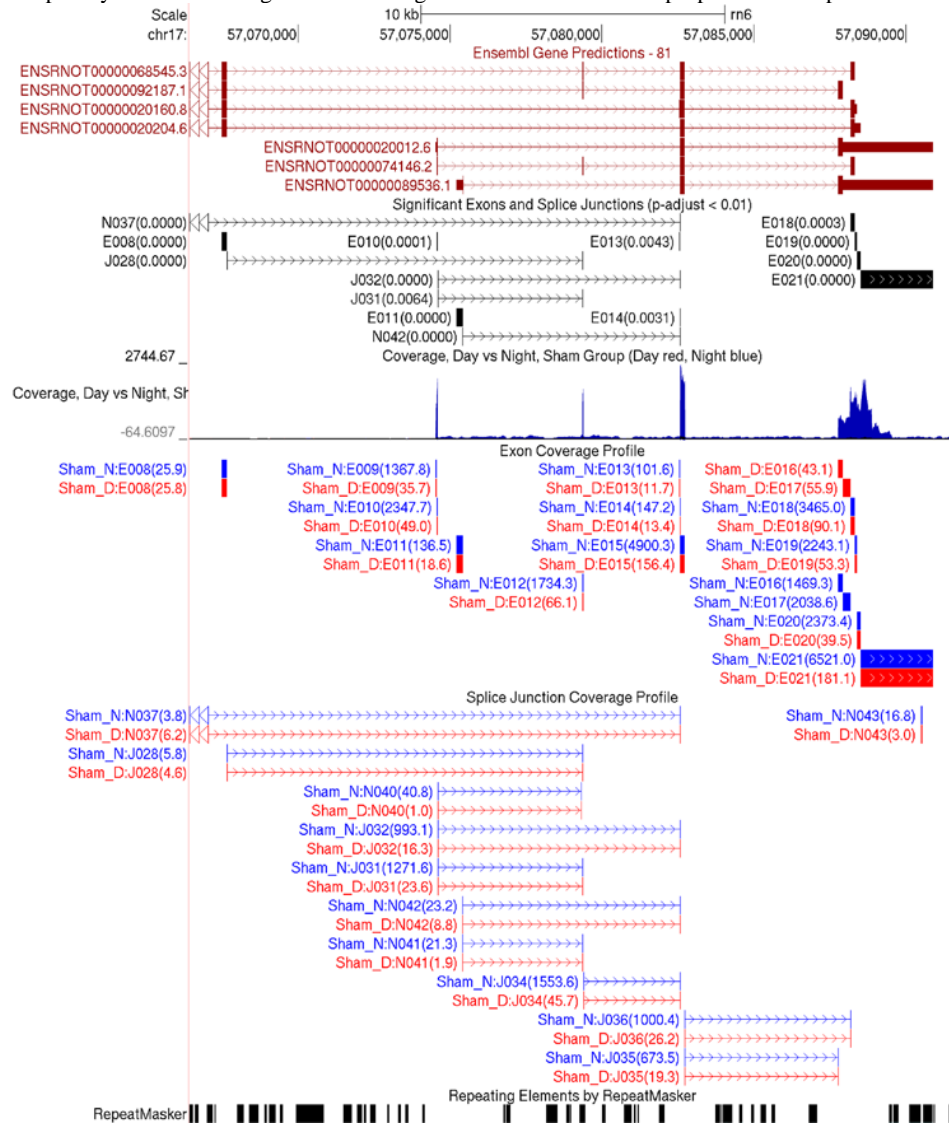

**Figure 4:** Day/Night gene profile plots for the Crem gene, created by *JunctionSeq* (a) and *DEXSeq* (b), both using an incomplete transcript annotation. These plots are equivalent to Figure **2** (a) and (b), except that all the transcripts except one (transcript ENSRNOT00000074146) were removed from the annotation prior to analysis. Without the *a priori* knowledge of the missing transcripts, DEXSeq cannot reliably detect differential usage. Note that the "novel" junction N010 is actually known junction J028 from Figure **2**. Similarly, N014 is J032 and N015 is J035. The other novel junctions are not present even in the full annotation. It should be noted that exon E001 shows borderline statistical significance in the *DEXSeq* plot (p-adjust = 0.016).

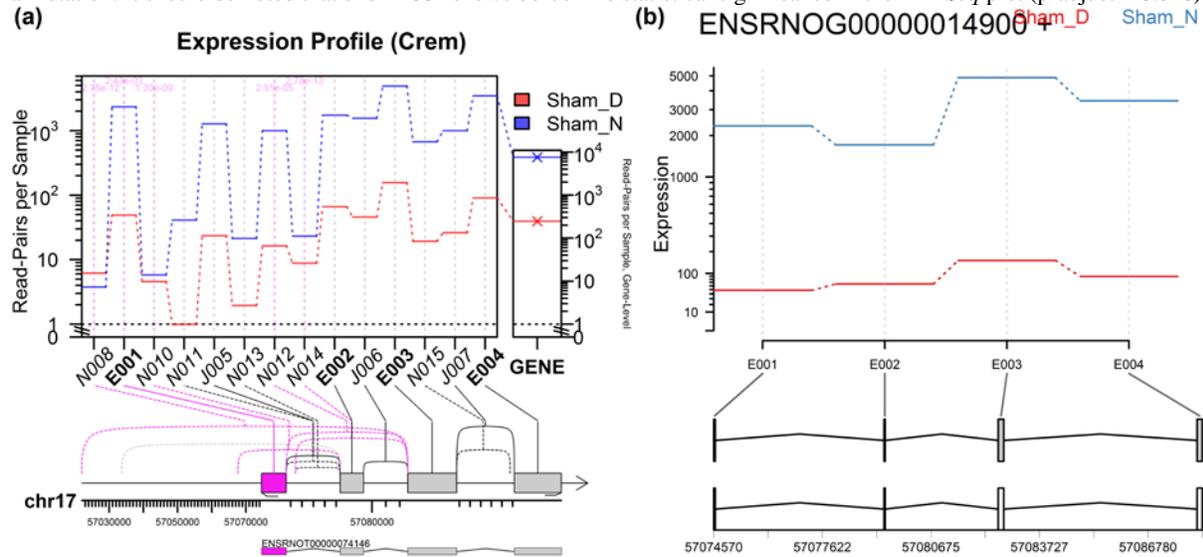

# Supplement

# 1 Supplementary Results

**Supplemental fig. 1: Gene profile plots, 4-hr vs control.**

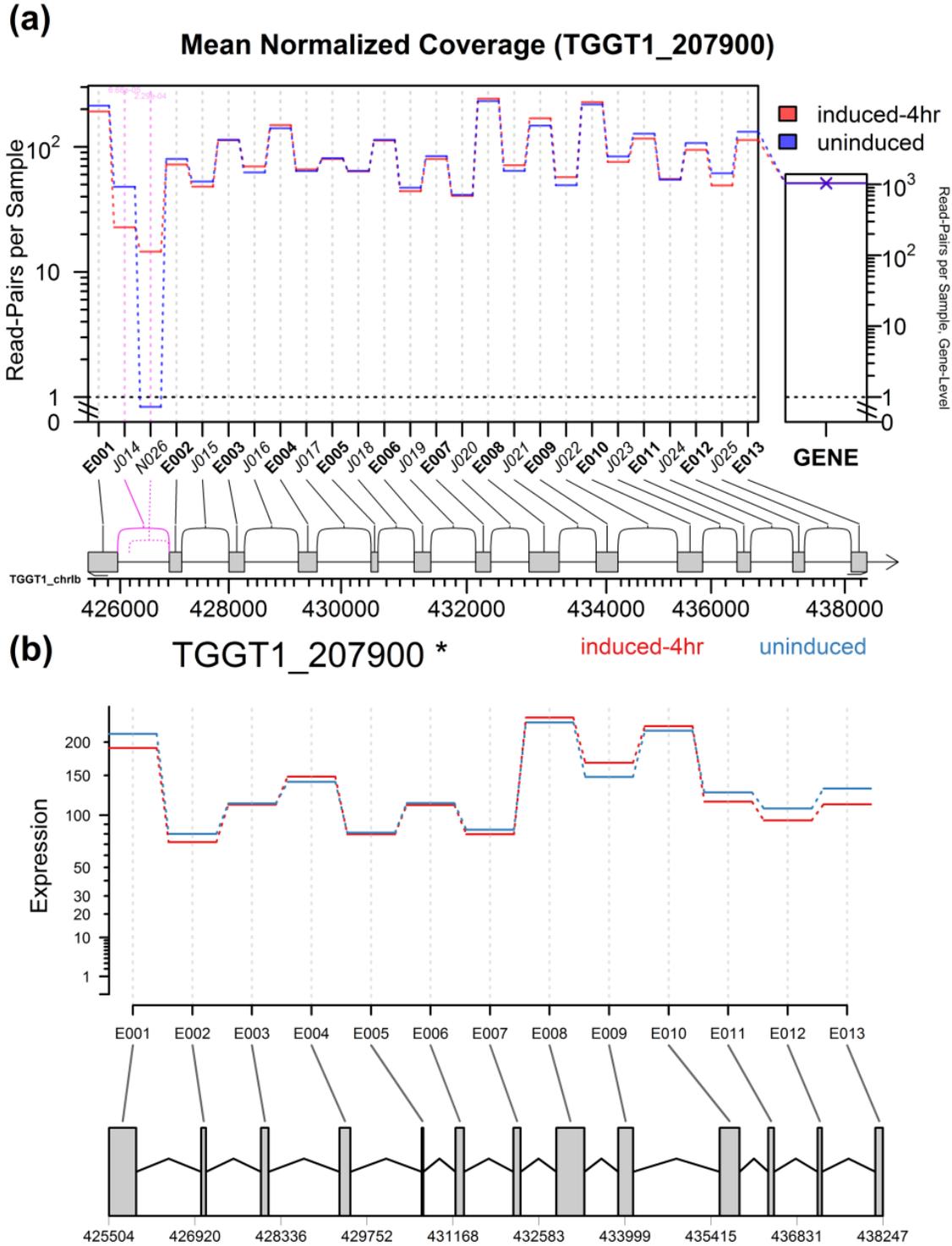

Displays the same information as figure 5 from the main text, except for the uninduced vs 4-hour-induced *Toxoplasma gondii* experiment.

**Supplemental fig. 2: Gene profile plots, 24-hr vs control.**

**(a)**

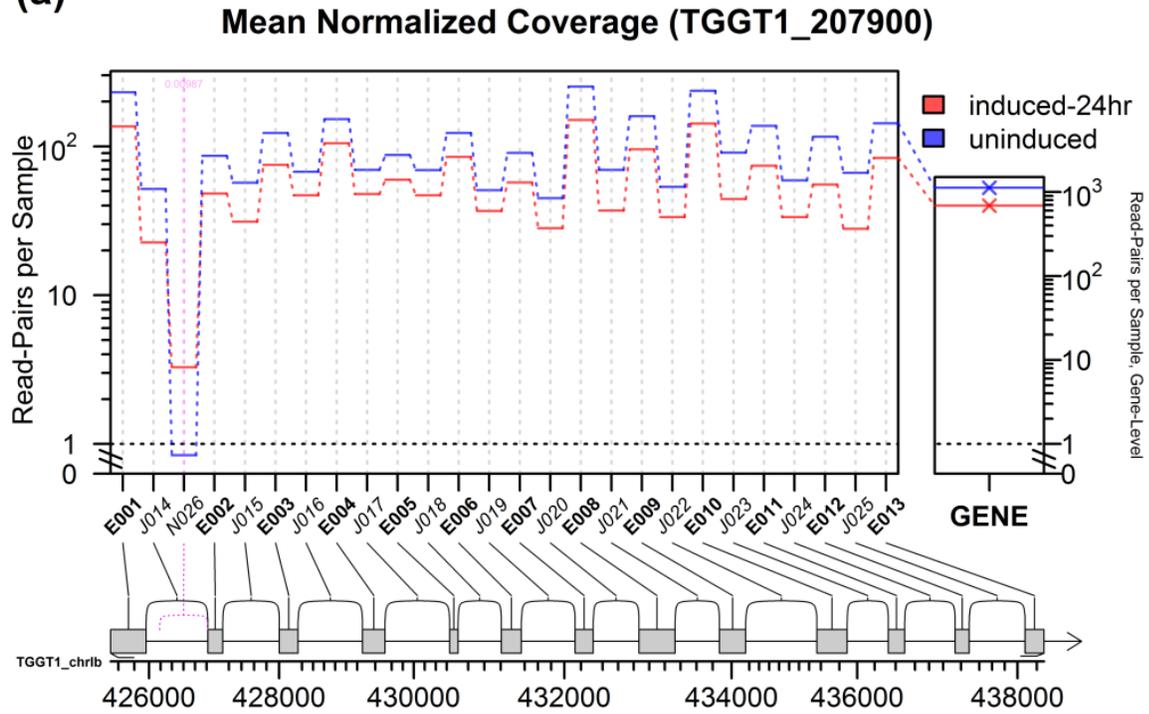

**(b)**

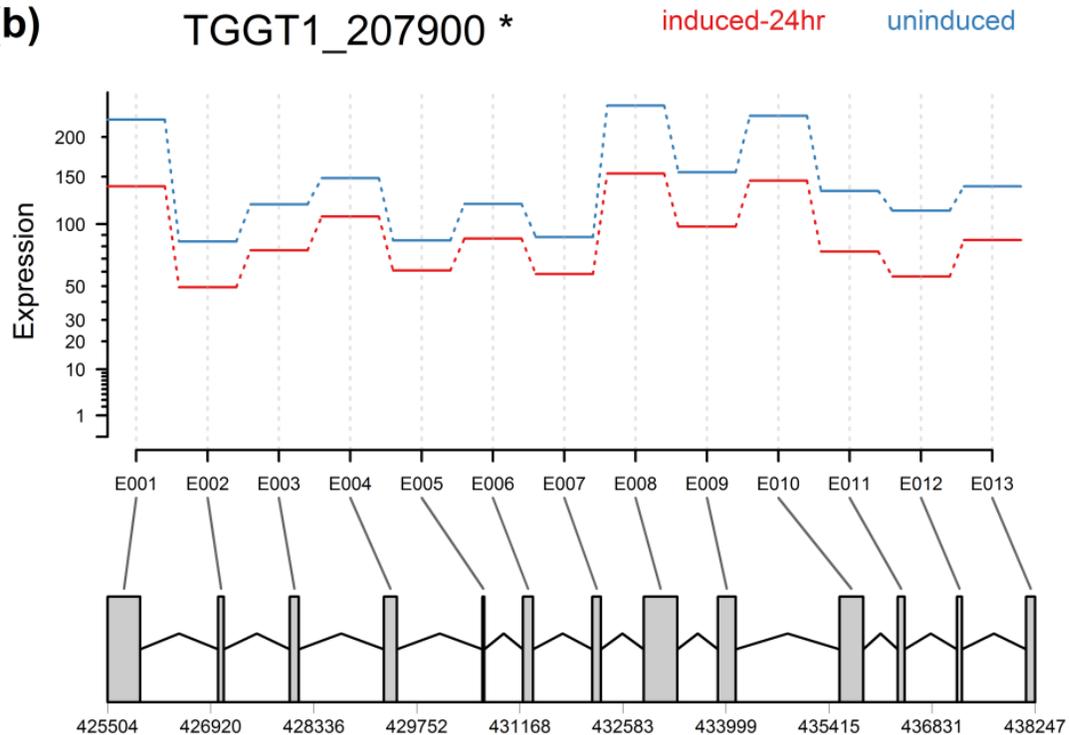

Displays the same information as figure 5 from the main text, except for the uninduced vs 24-hour-induced *Toxoplasma gondii* experiment.

**Supplemental fig. 3: JunctionSeq Results Venn diagrams.**

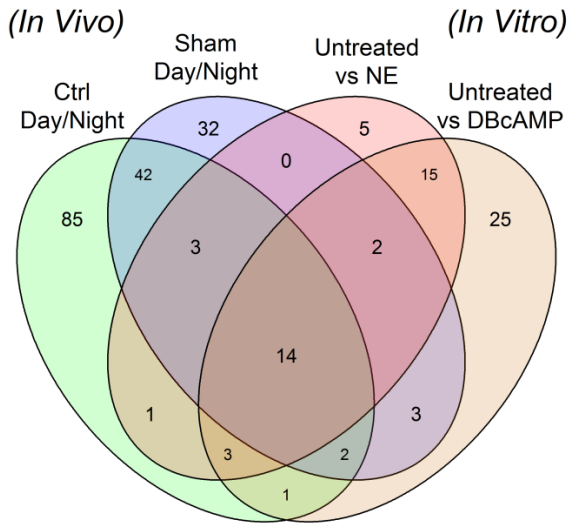

Venn diagrams describing the overlap in the genes found to possess splice junctions that display statistically significant (adjusted-p < 0.000001) differential usage in the four rat pineal analyses.

**Supplemental fig. 4: DEXSeq results Venn diagrams.**

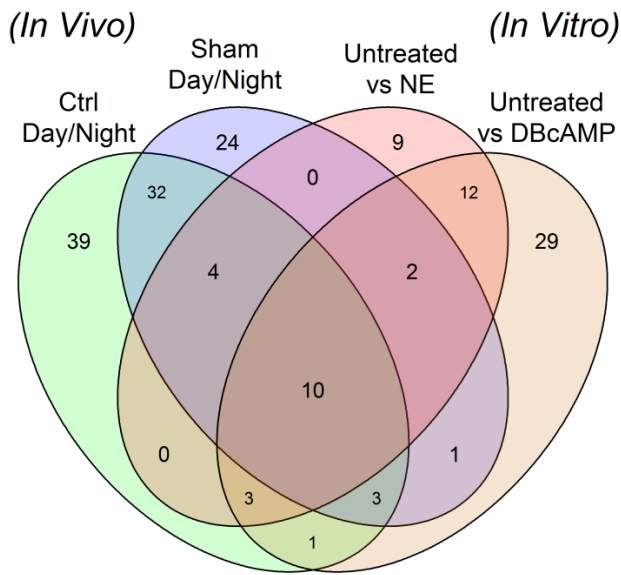

This Venn diagram, similar to the previous figure, displays the overlap between the results from the various in vivo and in vitro analyses performed by DEXSeq. As before, each cell displays the number of genes shared between the various analyses at adjusted-p < 0.000001.

**Supplemental Table 1: *DEXSeq* results.**

| Adjusted p-value threshold | *In vivo* | | | *In vitro* | | | Overlap, All Four |
|---|---|---|---|---|---|---|---|
| | Ctrl Day/Night | Sham Day/Night | Overlap, *in vivo* | CN vs NE | CN vs DBcAMP | Overlap, *in vitro* | |
| 0.01 | 289 | 240 | 110 | 121 | 192 | 86 | 28 |
| 0.001 | 189 | 152 | 85 | 90 | 129 | 67 | 20 |
| 0.0001 | 147 | 106 | 69 | 61 | 87 | 42 | 17 |
| 0.00001 | 112 | 82 | 56 | 44 | 67 | 32 | 10 |
| 0.000001 | 92 | 76 | 49 | 40 | 61 | 27 | 10 |

Similar to Table 1 from the main text. Lists the numbers of genes found by *DEXSeq* to exhibit significant differential exon usage for the five rat pineal gland analyses.

**Supplemental fig. 5: Gene profile plots for Crem, sham night/day.**

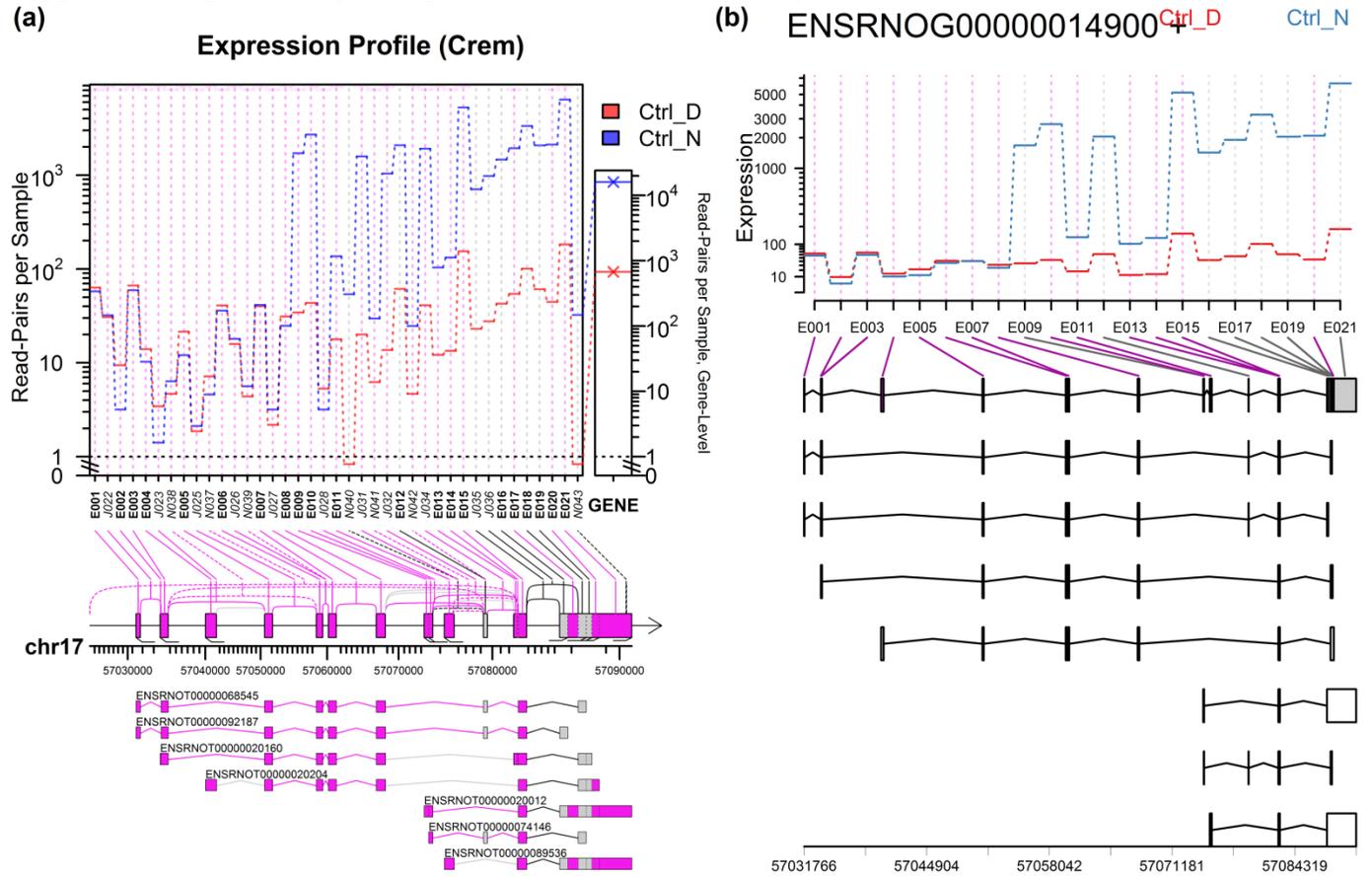

Comparison of gene profile plots generated by DEXSeq (a) and JunctionSeq (b) for the Crem gene in the sham day/night rat pineal gland experiment.

**Supplemental fig. 6: Gene profile plots for Crem, untreated vs norepinephrine.**

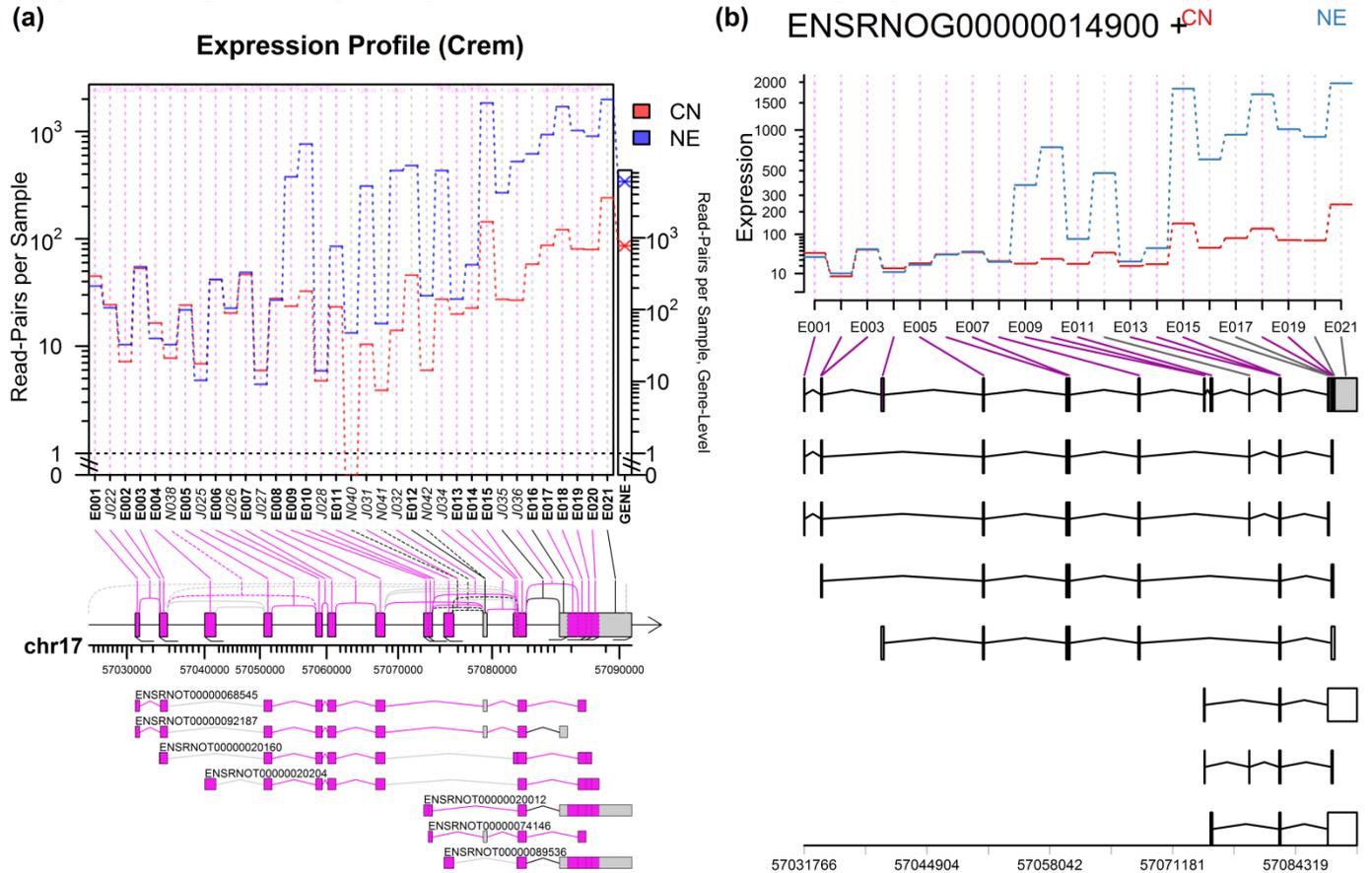

Comparison of gene profile plots generated by *JunctionSeq* (a) and *DEXSeq* (b) for the Crem gene in the untreated control vs DBcAMP-treated rat pineal gland experiment.

**Supplemental fig. 7: Gene profile plots for Crem, untreated vs DBcAMP.**

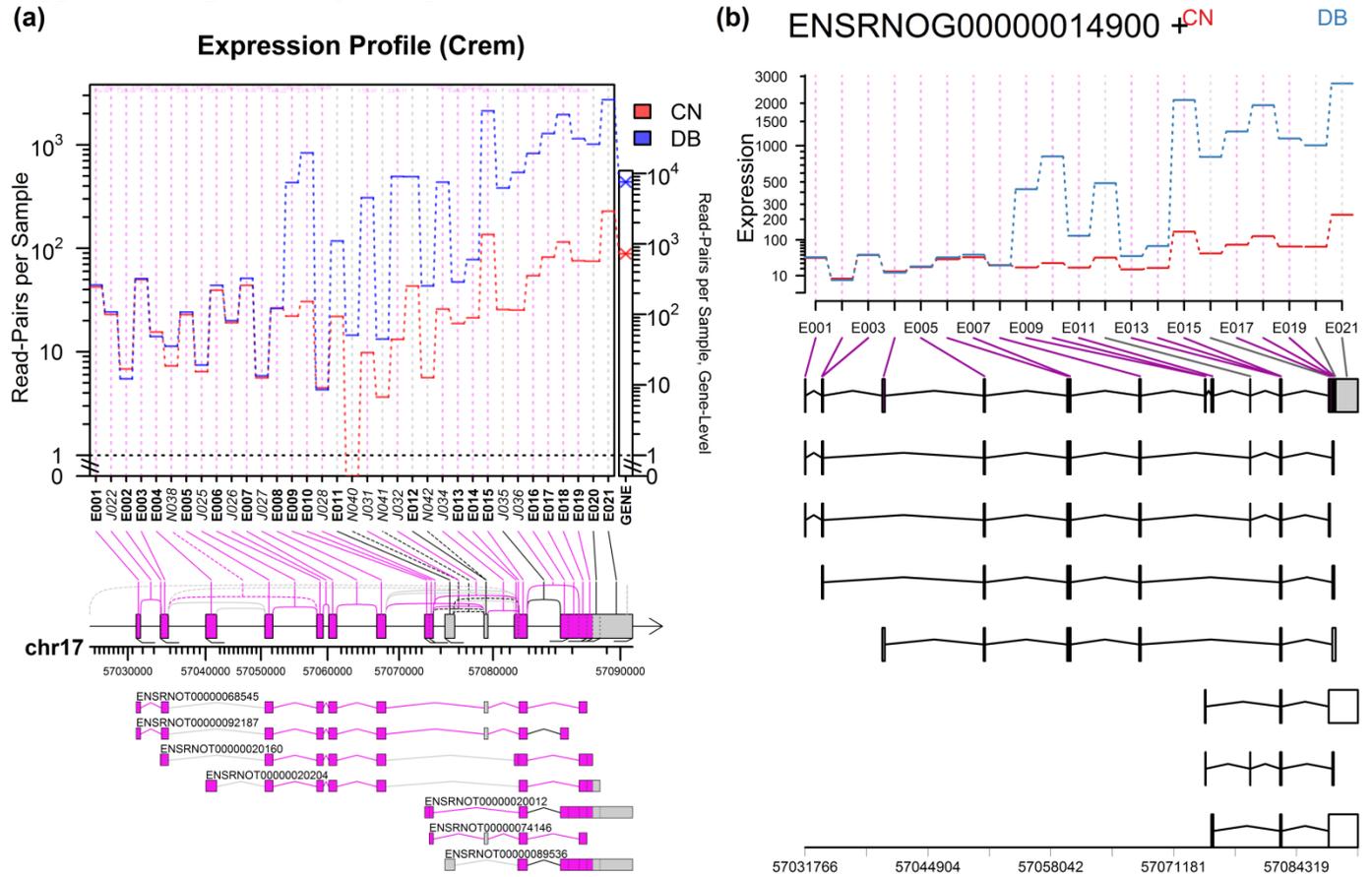

Comparison of gene profile plots generated by *JunctionSeq* (a) and *DEXSeq* (b) for the Crem gene in the untreated control vs NE-treated rat pineal gland experiment.

**Supplemental fig. 8: DEXSeq Gene profile plot, example human gene (simulated data).**

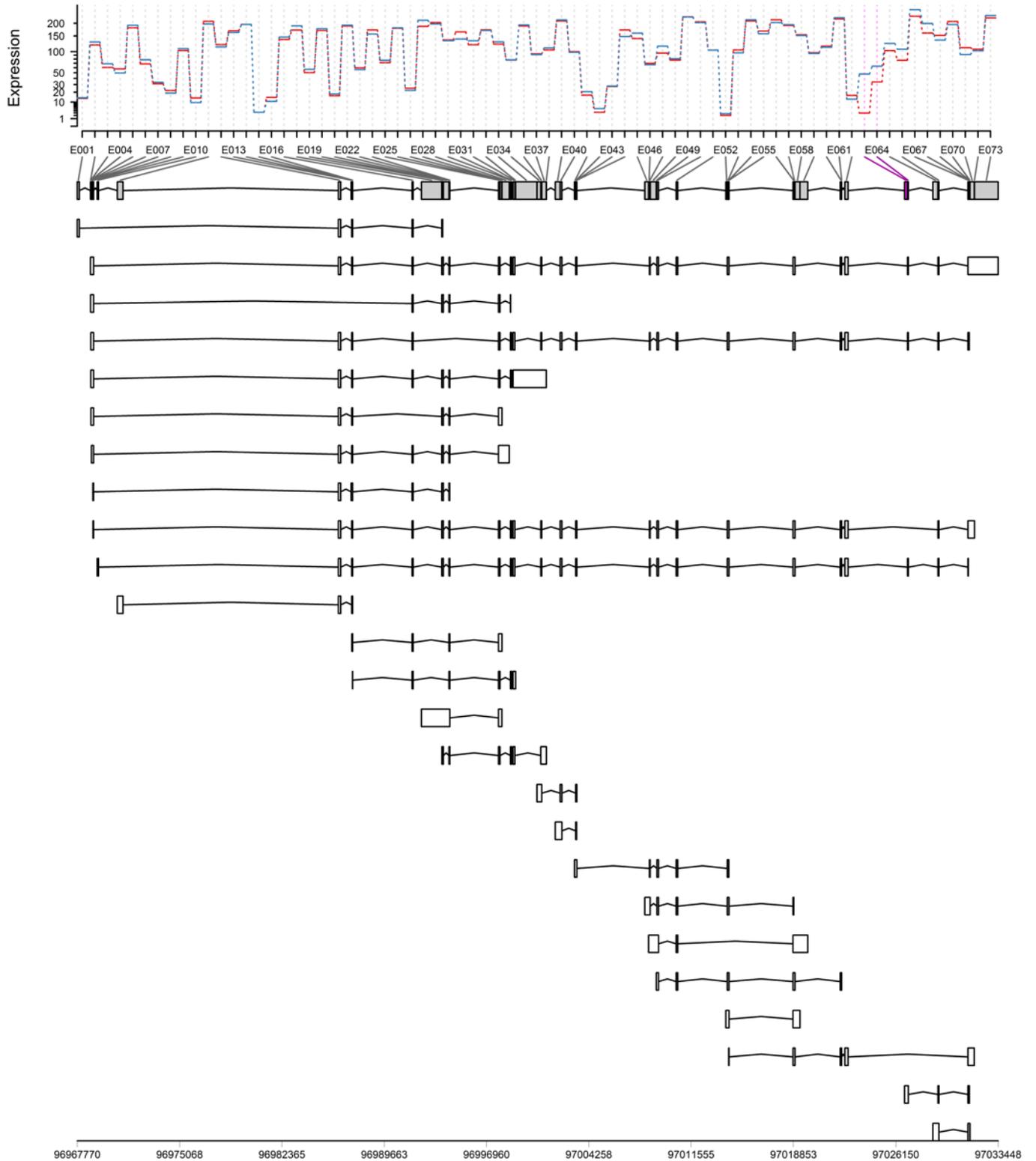

The above plot shows a DEXSeq plot for a moderately-complex human gene and a simulated dataset. Note that it is completely impossible to distinguish many exonic and splicing variants in the gene diagram, because many of the features are less than a pixel wide even at high resolution.

**Supplemental fig. 9: JunctionSeq Gene profile plot, example human gene (simulated data).**

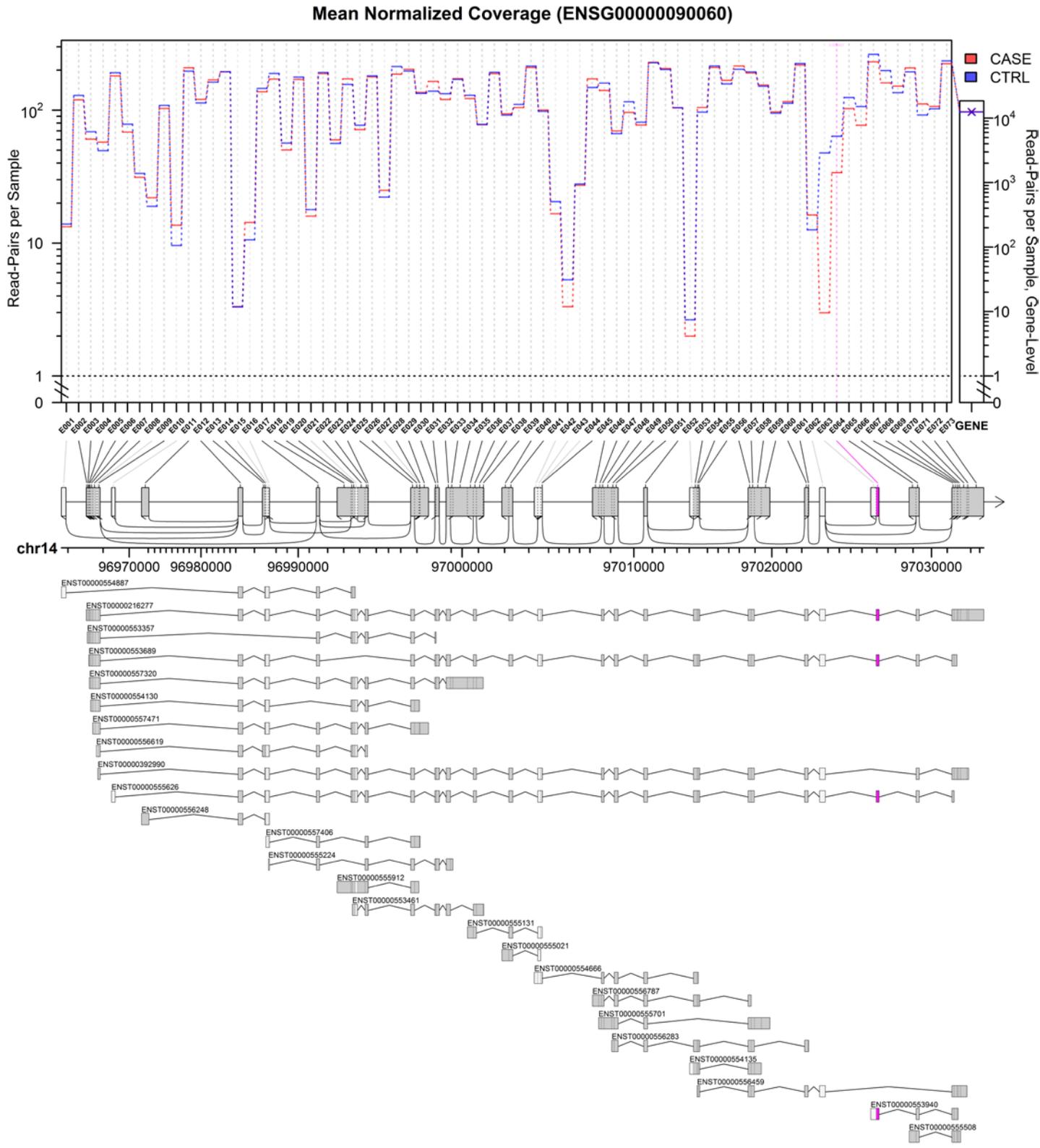

Using a specific set of optional parameters, JunctionSeq will reproduce a standard DEXSeq analysis, and then output plots using its own improved visualization engine. The above plot shows a JunctionSeq plot replicating the analysis that produced Supplemental fig. 8for a moderately-complex human gene and a simulated dataset. Note the various improvements in the lower plot produced by JunctionSeq. For example: exonic labels are shrunk and drawn at an angle, exonic and intronic regions are rescaled to improve readability, the overall gene-level expression is shown on the right, nested splice junctions are included, exon fragments are separated by dotted lines instead of solid ones, and exonic regions that do not satisfy the hypothesis test inclusion thresholds are drawn with a lighter gray.

# 2 Statistical Methodology

## 2.1 Model framework

Each exonic region or splice junction locus (or "feature") is fitted to a separate model. All terms from here forward are relative to a specific feature $j$, which is an exonic region or splice junction locus on gene $g$. Let there be $n$ biological replicates (or "samples").

For this feature $j$, we define two "counting bins": $\vec{y}_1 = (y_{11}, y_{12}, \ldots, y_{1n})$ and $\vec{y}_0 = (y_{01}, y_{02}, \ldots, y_{0n})$.

The counts $y_{1i}$ are defined as the number of reads (or read-pairs) in sample $i$ that cover the feature $j$. That is, the number of reads or read-pairs that intersect with the exonic region (if $j$ is an exon) or align over the splice junction (if $j$ is a splice junction). The counts $y_{0i}$ are defined as the number of reads (or read-pairs) that intersect with the gene $g$, but that do **not** cover the specific feature $j$. The gene-level counts are calculated using the standard HTSeq-based method used for DESeq2/edgeR differential gene expression.

Note: for the exon/junction counts, only read-pairs that actually align to the junction or exon itself are counted, read-pairs that only *flank* a feature are NOT counted towards that feature.

In the framework used by DEXSeq 1.12.2 (which, it should be noted, differs slightly from the original framework used by some earlier versions and presented in the DEXSeq methods paper) the feature counts ($y_{1i}$) are compared with the sum of all other feature counts belonging to the same gene. This means that some reads may be counted more than once if they span multiple features. When reads are relatively short (as was typical when DEXSeq was first introduced) this effect is minimal, but it becomes progressively less precise for longer reads. While, in theory, the methods used by DEXSeq should be robust against this issue, under certain circumstances it can result in unexpected artifacts. For example, if a gene has a large number of features in a very small genomic area (for example, if an exon has numerous alternative donor/acceptor sites), then reads covering that region may be disproportionately over-weighted, altering the relative fold change estimates by warping the linear contrasts.

Under our framework, no read-pair is ever counted more than once in each model.

These counts can be generated via QoRTs, a freely-available and open-source software package that provides both QC and data processing for RNA-Seq datasets (54).

The gene and exon counts could theoretically be generated by HTSeq, but there is currently no HTSeq-based method of generating the JunctionSeq input files which also include counts for known and novel splice junctions.

## 2.2 Novel splice junctions

In addition to the known splice junctions provided by a transcript annotation, novel splice junctions detected by the alignment software can be added to the analysis. These unannotated junctions can be selected for inclusion in the analysis if they had at least one endpoint within the span of a single gene and if they had a mean-normalized read-pair coverage (across all conditions) of greater than an assigned threshold (the default is 6 read-pairs per biological replicate).

## 2.3 Statistical Model

As in DEXSeq, we assume that the count $y_{bi}$ is a realization of a negative-binomial random variable $Y_{bi}$:

$$Y_{bi} \sim NB\big(mean = s_i\mu_{bi}, dispersion = \alpha_j\big)$$

Where $\alpha_j$ is the dispersion parameter for feature $j$, $s_i$ is the normalization size factor for sample $i$, and $\mu_{bi}$ is the normalized mean for sample $i$ and counting-bin $b$. (Remember that ($b = 1$) refers to the feature (ie. exon/splice-junction) counts and ($b = 0$) refers to the gene counts minus the feature counts.)

The normalization size factors $s_i$ are estimated using the "geometric" normalization method, which is the default method used by DESeq, DESeq2, DEXSeq, and CuffDiff (14,22,55,56). By default these normalizations are performed based on the gene-level counts.

## 2.4   Dispersion estimation

In many high-throughput sequencing experiments there are too few replicates to directly estimate the locus-specific dispersion term $\alpha_j$ for each feature $j$. This problem is well-characterized, and a number of different solutions have been proposed, the vast majority of which involve sharing information between loci across the genome (55,57). JunctionSeq uses the same method used by the DESeq2 package and by the more recent releases of the DEXSeq package. This method is described in detail elsewhere (23).

Briefly: individual-feature estimates of dispersion are generated via a Cox-Reid-based method. Then, a parametric model is fitted to these dispersions:

$$\alpha(\mu) = \frac{\alpha_1}{\mu} + \alpha_0$$

Where $\mu$ is the "base mean", or the sum of the normalized counts across the feature. The final dispersions are based on the maximum *a posteriori* (MAP) estimator, which combines information from the feature-specific and fitted dispersion estimates. The JunctionSeq R package also implements a number of other optional methods for estimating the dispersion (see Additional File 1).

## 2.5   Hypothesis tests for differential usage (DU)

The hypothesis test is performed using the same DESeq2-based methods used in DEXSeq v1.14.0, which are described in detail elsewhere (23). Note that the methods used in (23) include numerous improvements on the methods originally described in the DEXSeq methodology paper (22).

Briefly: two models are fitted to the mean $\mu_{bi}$. First, the reduced (null hypothesis) model:

$$\log(\mu_{bi}) = \beta + \beta_b^B + \beta_i^S$$

And then the alternative model:

$$\log(\mu_{bi}) = \beta + \beta_b^B + \beta_i^S + \beta_{\rho_i b}^{CB}$$

Where $\rho_i$ is the experimental condition value for sample $i$.

Note that the experimental-condition/counting-bin interaction term ($\beta_{\rho_i b}^{CB}$) is included, but the experimental-condition main-effects term ($\beta_{\rho_i}^{C}$) is absent. This term can be omitted because JunctionSeq is not designed to detect or assess gene-level differential expression. Thus there are two components that can be treated as "noise": variation in junction-level expression and variation in gene-level expression. As proposed by Anders et. al. (22), we use a main-effects term for the sample ID ($\beta_i^S$), which subsumes the condition main-effect term. This subsumes both differential and random variation (noise) in the gene-level expression, improving the power for detecting differential interaction between the count-bin term and the experimental-condition term.

## 2.6    Parameter Estimation:

While the statistical model described above is robust, efficient, and powerful, it lacks main effect terms and thus cannot be effectively used to estimate the size of the differential effect.

For the purposes of estimating the effect sizes and expression levels we create a separate set of generalized linear models for each feature. In this we diverge substantially from the current DEXSeq methods.

The mean $\mu_{bi}$ is modeled as:

$$\log(\mu_{bi}) = \beta + \beta_b^B + \beta_{\rho_i}^C + \beta_{\rho_i b}^{CB}$$

This model is used to calculate the parameter estimates $\hat{\beta}$, $\hat{\beta}_b^B$, $\hat{\beta}_{\rho_i}^C$, and $\hat{\beta}_{\rho_i b}^{CB}$, which are then used to calculate mean normalized coverage estimates $\hat{\mu}_\rho$ for each condition value $\rho$. Using linear contrasts, relative expression estimates can also be calculated for each junction locus and each condition value, producing an estimate of relative fold-change. This differs from the DEXSeq methodology, which fits one large model to the full set of features belonging to the given gene.

## 2.7    Multivariate Models:

If needed, additional covariates can be integrated into the generalized linear models. If we define $\tau_i$ as the covariate category for sample $i$, then we can define the hypothesis test reduced model as:

$$\log(\mu_{bi}) = \beta + \beta_b^B + \beta_i^S + \beta_{\tau_i b}^{TB}$$

And the alternative hypothesis model:

$$\log(\mu_{bi}) = \beta + \beta_b^B + \beta_i^S + \beta_{\tau_i b}^{TB} + \beta_{\rho_i b}^{CB}$$

Like the main effects term for the condition variable, the main effects term for the confounding variable is absent from both of the models used in the hypothesis test. This main effects term is subsumed into the sample-ID term $\beta_i^S$.

The parameter-estimation model can be extended similarly:

$$\log(\mu_{bi}) = \beta + \beta_b^B + \beta_{\rho_i}^C + \beta_{\tau_i}^T + \beta_{\tau_i b}^{TB} + \beta_{\rho_i b}^{CB}$$

In general the expression estimates are generated by averaging over confounding variable status. Optionally the expression estimates could be generated for each confounder status separately.

# 3   Test Dataset Data Processing and Methods

### 3.1   Application 1: *Toxoplasma gondii*

Our first test dataset originated from a previous study in which alternative splicing was detected and validated in *Toxoplasma gondii* between control samples and samples in which overexpression of the TgSR3 gene was induced (8). There were four sample groups of 3 biological replicates each: untreated; induced, 4 hours; induced, 8 hours; and induced, 24 hours. The dataset is available on the NCBI short read archive (SRA), accession number PRJNA252680.

Reads were realigned with *RNA-STAR* (9) to the ToxoDB v25 Toxoplasma gondii GT1 genome build and annotation (10). The dataset was processed via the *QoRTs* (1) data processing package and analyzed with *JunctionSeq*. Unlike with the previous study's analysis, a full *CuffLinks* transcript assembly was unnecessary as *JunctionSeq* can test novel splice junctions of known genes. We ran three analyses comparing each of the three induced groups to the untreated group.

## 3.2   Application 2: Circadian Rhythms in the Rat Pineal Gland

Our second test dataset consisted of 7 sample groups, 4 taken from live *Rattus norvegicus* pineal glands and 3 taken from rat pineal glands in organ culture, with 3 biological replicates each (21 biological replicate, total). The dataset is available online on the NCBI short read archive (SRA), accession number PRJNA267246. The surgical methods, sample collection, and sequencing is described in detail elsewhere (11). Briefly: the 12 *in vivo* samples were taken from no-surgery (Ctrl) and sham-surgery (Sham) rats at two time points: night and day. The 9 *in vitro* samples consisted of pineal glands in organ culture treated with norepinephrine (NE) or dibutyryl cyclic AMP (DBcAMP, an analogue of the second messenger, cAMP), as well as untreated controls (CN).

The 21 samples were aligned to the rn6 rat genome build using the *RNA-STAR* aligner (9) with the ensembl transcript annotation (release 80). Gene, exon, and splice-junction read-pair counts were then generated by the *QoRTs* data processing utility (1), adding novel splice junctions if they could be matched to a single known gene and if their mean normalized coverage across all samples exceeded 3. The five analyses were then carried out using *JunctionSeq* and *DEXSeq* using the standard developer-recommended options. The adjusted-p-value threshold for all plots was set to p-adjust < 0.01, which is the default for *JunctionSeq*.

### 3.2.1   Incomplete Annotation

To demonstrate *JunctionSeq's* ability to detect differential usage in novel splice junctions even with an incomplete transcript assembly, we performed a second set of analyses with a reduced annotation. For each of the four known AIR genes (Crem, Pde4b, Slc15a1, and Atp7b), we manually removed all but one transcript from the ensembl annotation GTF and then re-ran the analyses. Since the gene Slc15a1 only has one known transcript, it was not altered in this analysis.

In general it was not possible to uniquely identify the dominant isoform from each gene. When selecting which isoform to leave in the annotation, we chose one of the most likely dominant isoforms based on the exon and junction coverages found by *JunctionSeq*. In truth it does not affect the validity of the experiment, since in practice the annotated isoform may be any one of the true isoforms (or in some cases even be a false isoform that does not actually exist). See Supplemental Table 2 for a full listing of the isoforms used in the partial annotation analysis.

All genes/transcripts other than the ones belonging to the four known AIR genes were included in the analysis as normal. They had to be included because *JunctionSeq* shares information across genes in order to estimate the size factors and biological dispersion.

*Supplemental Table 2: Genes and transcripts selected for use in the "incomplete annotation" analysis.*

| Gene Symbol | # Known Transcripts | Gene Ensembl ID | Transcript used in the "incomplete" annotation |
|---|---|---|---|
| Atp7b | 2 | ENSRNOG00000012878 | ENSRNOT00000089265 |
| Crem | 7 | ENSRNOG00000014900 | ENSRNOT00000074146 |
| Pde4b | 3 | ENSRNOG00000005905 | ENSRNOT00000007738 |
| Slc15a1 | 1 | ENSRNOG00000011598 | ENSRNOT00000015890 |

It should be noted that the adjusted-p-values listed in the "full" and the "incomplete" analyses for the gene Slc15a1 were slightly different, despite the fact that the gene annotations were unchanged for this gene. This was due to (very slight) analysis-wide differences in the two sets of analyses. *JunctionSeq* and *DEXSeq* both share information between genes to calculate size factors and dispersion estimates. Furthermore, the FDR-based Benjamini-Hochberg multiplicity correction procedure causes the adjusted-p-values of each test to depend on the other p-values in each set (12). Thus, changes to one gene can (usually only slightly) change the results for the other genes.